\documentclass[12pt,a4paper]{iopart}
 
\usepackage{graphicx}
\begin{document}

\title{Continuous-time cross-phase modulation and quantum computation}

\author{Jeffrey H Shapiro and Mohsen Razavi}

\address{Research Laboratory of Electronics, Massachusetts Institute of Technology, Cambridge, MA 02139, USA}
\ead{jhs@mit.edu and mora158@mit.edu}
\begin{abstract}
The weak nonlinear Kerr interaction between single photons and intense laser fields has been recently proposed as a basis for distributed optics-based  solutions to few-qubit applications in quantum communication and computation. Here, we analyze the above Kerr interaction by employing  a continuous-time multi-mode model for the input/output fields to/from the nonlinear medium. In contrast to previous single-mode treatments of this problem, our analysis takes into account the full temporal content of the free-field input beams as well as the non-instantaneous response of the medium. The main implication of this model, in which the cross-Kerr phase shift  on one input is proportional to the photon flux of the other input, is the existence of phase noise terms at the output. We show that these phase noise terms will preclude satisfactory performance of the parity gate proposed by Munro, Nemoto, and Spiller \cite{weak}. 
\end{abstract}

\maketitle

\section{Introduction}

Years of experimental and theoretical research on quantum information science have revealed the numerous difficulties that must be overcome to build a large-scale quantum computer. Such a computer could factor large numbers and search unstructured databases far more efficiently than any classical computer. To pave the road to quantum computer implementations, we first need to  master few-qubit technology. One appealing approach to the latter problem uses optics-based configurations for quantum processors.  Light can carry quantum information in the form of single-photon polarization states, or vacuum plus single-photon superpositions.  Spontaneous parametric down-conversion can produce entangled photon pairs and heralded single photons.  Beam splitters and wave plates can be used to convert between single-photon polarization and vacuum plus single-photon qubits.  They can also accomplish arbitrary single-qubit rotations.  To complete a universal gate set for quantum computation, all that is then needed is an appropriate  two-qubit quantum gate, such as a controlled-{\sc not} ({\sc cnot}) gate. A universal gate set will also serve quantum communication systems, in that it can realize their required full Bell-state measurement (BSM) device. 

Realizing an all-optical {\sc cnot} requires a nonlinear interaction between single photons, but traditional nonlinear materials offer only weak coupling between single photons.  Linear optics quantum computing \cite{KLM, Pittman} circumvents this difficulty by using photodetection and post-selection to provide a strong nonlinearity at the cost of a non-deterministic architecture that needs substantial ancillary resources.  Electromagnetically-induced transparency \cite{schmidt} may afford a strong cross-Kerr effect at the single-photon level, and lead to a deterministic all-optical {\sc cnot} implementation.  Munro, Nemoto, and Spiller \cite{weak}, however, have made a rather different suggestion for an all-optical {\sc cnot}.  By successive \em weak\/\rm\ cross-Kerr interactions between a strong coherent-state probe beam---that acts as a quantum communication bus---and a pair of single-photon qubit beams they realize a deterministic parity gate from which a {\sc cnot} can be constructed.   Their parity gate can also be used to realize a full BSM apparatus.  Munro, Nemoto, and Spiller use a single-mode treatment of cross-phase modulation (XPM) to show that their parity gate achieves near-ideal performance.   It has been known for some time, however, that a proper quantum theory for self-phase modulation (SPM) requires a  continuous-time multi-mode theory \cite{haus}. Moreover, we have recently extended the continuous-time quantum theory of SPM to the case of XPM  \cite{jeff_kerr}.  Using that theory we proved that  the phase noises needed to ensure preservation of the free-field commutator brackets at the output of the XPM interaction precluded high-fidelity operation of the quantum phase gate proposed by Chuang and Yamamoto \cite{ike}, which relies on a {\em strong} nonlinear interaction between two single photons in a cross-Kerr medium.  The question that we will address in the present paper is whether a similar fidelity degradation will occur in the parity gate from \cite{weak} when its performance is assessed in the continuous-time framework.
 
There are two issues that necessitate employing a continuous-time model for the cross-Kerr effect. First, the photonic qubits at the input to and output from a quantum gate will be flying qubits, i.e., free-field optical pulses.  Single-mode treatments of the quantum gate that ignore the temporal behavior of the optical pulses have assumed---either explicitly or implicitly---that gate operation occurs within a high-finesse cavity. However, even then free-field inputs and outputs must be properly accounted for so that the fidelity of a multi-gate quantum circuit can be determined.  In other words, if we put our optical gate in a black box, then its input-output relation is a transformation from a continuous-time free-field operator at its input to another continuous-time free-field operator at its output. 

The second issue that requires a multi-mode treatment of XPM, is the non-instantaneous nature of the cross-Kerr interaction.  Optical fibers have a 5--10~fs response time for the SPM interaction \cite{haus}, and a similar response time is expected for XPM.  The response time of the nonlinear material plays a key role in its quantum behavior when single-photon pulses are involved.  For example, we have shown \cite{jeff_kerr} that if the XPM response time is much shorter than the time duration of a pair of single-photon input pulses, then essentially no XPM phase shift will occur on these pulses even when the peak nonlinear phase shift from a single photon is the $\pi$\,rad that Chuang and Yamamoto require for their single-photon controlled-phase gate \cite{ike}.  This counter-intuitive behavior occurs because the $\pi$-rad phase shift only affects a very short---and randomly-distributed in time---portion of the single-photon pulse.  Conversely, in the slow-response regime---wherein the optical pulses are much shorter than the XPM response time---we can adjust the timing of the input single-photon pulses so that one photon can induce an appreciable XPM phase shift on the other.

For an XPM medium whose response function is causal and non-instantaneous, preservation of the free-field commutator relations for the output field operators imposes phase noise terms on these outputs. These phase noises will be seen, in this paper, to severely degrade the fidelity of the parity gate proposed by Munro, Nemoto, and Spiller.  It will also be shown that the mean-squared phase noise is proportional to the response function's amplitude, implying that stronger nonlinearity is accompanied by increased phase noise.
 
In this paper, we first summarize the model introduced in \cite{jeff_kerr}  for the XPM interaction, which applies in the absence of loss, dispersion, and SPM.  Then, in Section~\ref{Sec5coh}, we apply the slow-response version of this model to a simple gate in which a single-photon pulse induces a weak phase shift on a coherent-state probe beam. This gate is the building block for the Munro, Nemoto, and Spiller  parity gate, which will be described  and analyzed in Section~\ref{Sec5parity}. That treatment constitutes the continuous-time fidelity analysis of their parity gate.  Section~\ref{Sec5numeric} includes some numerical results, which quantify the parity gate's phase-noise induced fidelity loss.  Section~\ref{Sec5conc} concludes the paper by discussing the applicability of our results to different scenarios.

\section{Continuous-time cross-phase modulation}
\label{Sec5model}

In this section, we describe the continuous-time model for XPM, introduced in \cite{jeff_kerr}. This model applies primarily to the Kerr interactions that occur in a length of a macroscopic material, e.g., an optical fiber. Hence, its input and output field operators are free fields.  Whereas discrete modes are proper choices for modeling light that is confined in a cavity, a continuous-time formalism is more appropriate for free-field operators. In this formalism, 
a positive-frequency photon-units field operator associated with a $+z$-going  electric field, in a well-defined polarization, can be written as follows \cite{yuenjeff1, Loudon}
\begin{equation}
\label{fieldop}
\hat E(t) =  \int\! {\frac {{\rm d}\omega}{\sqrt{2 \pi}}\, \hat a(\omega) e^{-i\omega t}},
\end{equation}
where  $\hat a(\omega)$ is the annihilation operator associated with frequency $\omega$ that satisfies $[\hat a(\omega), \hat a^\dag (\omega')] = \delta (\omega - \omega')$. The limits of integration in  (\ref{fieldop}) are from $0$ to $\infty$. However, for the optical sources we shall consider, whose bandwidths are narrow in comparison to their common center frequency $\omega_0 \gg 0$, we can extend the integral's lower limit to $-\infty$.   It then follows that
\begin{equation}
[\hat E(t),\hat E^\dag(t')] = \delta (t-t')
\end{equation}
is the free-field commutator.  

For our later calculations it will be useful to write our field operators in terms of a discrete set of basis functions:
\begin{equation}
\label{ctod}
\hat E(t) = \sum_i {\hat a_i\phi_i(t)},
\end{equation}
where $\{\phi_i\}$ is a complete orthonormal set of functions satisfying
\begin{equation}
 \int\! {{\rm d} t\, \phi_i(t) \phi_j^\ast (t)} = \delta_{ij} {\rm \ \ and \ \ }
 \sum_i {\phi_i^\ast(t) \phi_i (t')} = \delta(t-t'),
\end{equation}
and $\hat a_i = \int\! {{\rm d} t\, \phi_i^\ast (t) \hat E(t)}$ is a discrete-mode annihilation operator, which satisfies $[\hat a_i, \hat a_j^\dag ] = \delta _{ij}$. Equation (\ref{ctod}) provides us with a prescription for converting a continuous-time field operator to a sum of discrete-mode operators, where each mode has a pulse shape orthogonal to that of the other modes. We will use this formalism frequently in forthcoming sections.  Strictly speaking, such a discrete representation should be applied to a time-limited or band-limited field, but inasmuch as we will only need one mode per field for our study of slow-response XPM, no loss of generality is entailed by use of (\ref{ctod}).

\begin{figure}
\centering
\includegraphics [height=2cm] {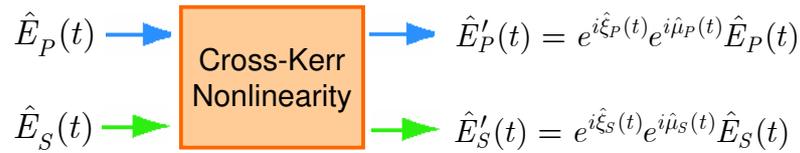}
\caption{\label{Shapiro_Razavi_fig1} Modeling the cross-Kerr nonlinearity using input/output field operators. Here, $\hat \xi_S$ and $\hat \xi_P$ are phase-noise operators, while $\hat \mu_{S}$ and $\hat{\mu}_{P}$ are XPM operators that depend on the probe-beam and signal-beam photon-flux operators, respectively, as well as the medium's response function.}
\end{figure}

Now, let us begin our consideration of the Kerr medium. We will assume that there is no loss, no dispersion, and no SPM in this medium. Furthermore, we denote the photon-units input field operators by $\hat E_S(t)$ for the signal beam and $\hat E_P(t)$ for the probe beam, whose respective photon-units output field operators are $\hat E'_S(t)$ and $\hat E'_P(t)$; see figure~\ref{Shapiro_Razavi_fig1}. Then, the input-output relationship for this material is given by \cite{jeff_kerr}
\begin{equation}
\label{output_kerr}
\hat E'_K(t) = e^{i\hat \xi_K(t)} e^{i\hat \mu_K(t)} \hat E_K(t), \quad\mbox{for $K=S,P$,}
\end{equation}
where we have suppressed the group delay, and
\begin{equation}
\label{kerr_effect}
 \hat \mu_K(t) \equiv \kappa \int\! {{\rm d} \tau\, h(t-\tau) \hat E^\dag_J(\tau) \hat E_J(\tau)} , \quad\mbox{for $J,K \in \{S,P\}, J \neq K$}
\end{equation}
gives the XPM phase shifts on the signal and the probe beams. Here, $\kappa$ is the XPM coupling coefficient, and $h(t)$ is a causal response function, which has been normalized to satisfy $\int\! {{\rm d}t\, h(t)} =1$. Equation (\ref{kerr_effect}) is in accord with our semiclassical understanding of the Kerr effect, in which we assume that the nonlinear operators responsible for the Kerr effect are proportional to the photon-flux operators. This is what we expect to hold for an ideal cross-Kerr medium. However, for the output-field operators to commute with each other, i.e.,  
\begin{equation}
\label{output_commut}
[\hat E'_S(t),\hat E'_P(t')] =  [\hat E'_S(t),\hat E'^\dag_P(t')] = 0,
\end{equation}
as is the case for their corresponding input-field operators, 
we need to include a pair of Langevin noise operators.  These noise operators represent coupling to localized noise oscillators that typically represent the molecular vibrations in the medium, and they result in Hermitian phase-noise operators $\hat \xi_S(t)$ and $\hat \xi_P(t)$ in the field-operator input-output relations (\ref{output_kerr}).  The coupling coefficient between the light and the corresponding reservoir mode at frequency $\Omega$ turns out to be proportional to $\sqrt{H_i(\Omega)}$, where
\begin{equation}
H_i(\Omega) = \int\!  {{\rm d} t \, h(t) \sin(\Omega t)}
\end{equation}
must be non-negative for $\Omega \geq 0$, so that all the damping coefficients are positive. Furthermore, equation~(\ref{output_commut}) requires that \cite{jeff_kerr}
\begin{equation}
[\hat \xi_S(t),\hat \xi_P(u)] = i\kappa [h(u-t) - h(t-u)].
\end{equation}
In thermal equilibrium, $\hat \xi_S(t)$ and $\hat \xi_P(t)$ can be taken to be in zero-mean joint Gaussian states with the following symmetrized correlation function
\begin{eqnarray}
\label{phase_corr}
\langle \hat \xi_K(t)\hat \xi_K(u) + \hat \xi_K(u) \hat \xi_K(t) \rangle =  \nonumber \\
\hspace{.2in}
\kappa \int\!{\frac {{\rm d} \Omega}{\pi}\,H_i(\Omega) \coth[\hbar \Omega/ (2 k_B T)]\cos[\Omega(t-u)]}, \quad\mbox{for $K=S,P$,}
\end{eqnarray}
where $k_B T$ is the thermal fluctuation energy.

The above model for XPM has several interesting implications. First, from equation~(\ref{phase_corr}), it is seen that the phase noise variance at time $t$, $\langle \hat \xi_K^2(t) \rangle $, for $K=S,P$, is nonzero even if $T \rightarrow 0$. This can result from photon-phonon interactions in the medium \cite{haus}. Second, it can be easily shown that an instantaneous response function in this model cannot reproduce the well-known classical results for XPM \cite{jeff_kerr}. The fact that the response function has a nonzero effective time duration $\Delta$ becomes important  when we are dealing with single-photon pulse inputs, because the nonlinear phase shift imparted by a single-photon pulse depends on its duration being greater or smaller than $\Delta$ \cite{jeff_kerr}. Let us elaborate more on this issue by considering  the special case of a single-photon signal input in the slow-response regime.

A single-photon signal pulse can be represented by the following state
\begin{equation}
|1\rangle_S = \int\! {{\rm d} t\, \phi(t) |1_t\rangle_S},
\end{equation}
where the wave function $\phi(t)$ satisfies $\int\!{\rm d}t\,|\phi(t)|^2 = 1$.  
Here, $|1_t\rangle_S$ is a multi-mode state that represents a single photon at time $t$; it satisfies $\hat E_S(t') |1_t\rangle_S = \delta(t-t') |{\bf 0} \rangle_S$, where $|{\bf 0} \rangle_S$ is the multi-mode vacuum state for the signal beam. It follows that $|\phi(t)|^2$ is the probability density function for observing this photon at time $t$. Equivalently, we can use the  discrete formalism from equation~(\ref{ctod}) by employing a complete orthonormal basis in which $\phi_1(t) = \phi(t)$.  Using such a basis, we have that $|1\rangle_S$ is a state for which $\hat a_1$ is in the number state $|1\rangle_{a_1}$, while the other $\{\hat{a}_i\}$ are in their vacuum states.  The average XPM phase-shift factor introduced by this single photon on the probe beam is therefore 
\begin{equation}
\label{ph_sph}
\langle e^{i\hat \mu_P(t)} \rangle  =   \int\!{{\rm d} \tau\, |\phi(\tau)|^2 e^{i \kappa h (t-\tau)} } ,
\end{equation}
where we have used \cite{Loudon}
\begin{equation}
\label{mf}
\exp\left[\int\!{{\rm d} t\, g(t) \hat E^\dag(t) \hat E(t)}\right] = {\cal{N}} \left\{ \exp\left[\int\!{{\rm d} t\, (e^{g(t)}-1) \hat E^\dag(t) \hat E(t)}\right] \right \},
\end{equation}
with ${\cal{N}}\{f(\hat E^\dag, \hat E) \}$ denoting the normally-ordered form of the operator $f(\hat{E}^\dag, \hat{E})$. 

In the {\em slow-response} regime,  $|\phi(\tau)|^2$ behaves like $\delta(\tau-t_0)$, relative to $h(t-\tau)$, for an appropriate $t_0$ near the pulse's center.   In other words, the signal pulse's time duration, $\tau_0$, is much shorter than $\Delta$, the time duration of the response function.  From equation~(\ref{ph_sph}), we then obtain
\begin{equation}
\langle e^{i\hat \mu_P(t)} \rangle  =  e^{i \kappa h(t-t_0)}.
\end{equation}
Now, if the probe pulse shape\footnote{Interferometric measurements are crucial to optics-based quantum computation.  Thus we will assume that the probe and the signal beams have identical pulse shapes except for a timing offset $t'$. This assumption is not essential, however, for our current discussion of the slow-response XPM. All we really need is for the probe pulse's effective duration to be comparable to that of the signal pulse, and hence much shorter than that of the response function.} is $\phi(t-t')$, then the phase shift induced by our single photon on the probe's pulse at $t=t_0 + t'$ is $\kappa h(t')$. We can maximize this phase shift by choosing $t' = t_h$, where $h(t_h) = \max_t [h(t)]$.

Given that $\Delta \approx  {\rm 1-10~fs}$ for optical fiber, it is impractical to work in its slow-response regime. However, this is the only regime in which a useful quantum interaction may be seen. In the fast-response regime---in which, $\tau_0 \gg \Delta$---only a $\Delta$-duration portion of the probe pulse undergoes a nonlinear phase shift, and the time location of this region is randomly distributed over the entire probe pulse with probability density function $|\phi(t)|^2$ \cite{jeff_kerr}. This property of the fast-response regime precludes its being useful for optics-based quantum computation.
Therefore, we will limit our subsequent work to the slow-response regime.  If promising results are obtained, it will then behoove us to find a Kerr medium with a more useful response time\footnote{In the above analysis, we have assumed that there is no SPM in our nonlinear material. It turns out that even if the SPM effect is present in the medium, it can be suppressed by  operating in the slow-response regime. This SPM suppression occurs when $h(t_h) \gg h_{\rm SPM}(0)$, where $h_{\rm SPM}(t)$ is the SPM response function for the probe beam.}.

\section{Kerr nonlinearity between a single photon and a coherent state}
\label{Sec5coh}

Consider a Kerr medium with XPM coupling constant $\kappa$ and a response function $h(t)$, in which  there is no loss, no dispersion, and no SPM. Suppose we illuminate this medium with a signal pulse and a probe pulse, with the former being in a superposition of the vacuum state and the single-photon state with pulse shape $\phi(t+t_h)$, viz., 
\begin{equation}
|\psi_{\rm in}\rangle_S = \alpha |{\bf 0}\rangle_S + \beta \int\! {{\rm d} t\, \phi(t+t_h) |1_t\rangle_S} , 
\end{equation}
with $|\alpha|^2 + |\beta|^2 = 1$,
and the latter being in the coherent state $|\alpha_P \phi(t) \rangle$, for which $\hat E_P(t)|\alpha_P \phi(t) \rangle = \alpha_P \phi(t) |\alpha_P \phi(t) \rangle$. In the mode-decomposition formalism, we can think of the signal pulse as being in the superposition state $|\psi_S\rangle = \alpha|0\rangle_{a_S} + \beta |1\rangle_{a_S}$, where $|n\rangle_{a_S}$ is the Fock state associated with the modal annihilation operator $\hat a_S \equiv \int\! {{\rm d} t\, \phi^\ast(t+t_h) \hat E_S(t)}$. Similarly, the probe beam can be taken to be in a coherent state $|\psi_P\rangle = |\alpha_P\rangle_{a_P}$ associated with the modal annihilation operator $\hat a_P \equiv \int\! {{\rm d} t\, \phi^\ast(t) \hat E_P(t)}$. All other input modes are in their vacuum states. The time shift between the signal and probe pulses results in the signal's inducing the maximum nonlinear phase shift on the probe pulse  in the slow-response regime. The goal of this section is to determine the output density operator for this simple gate, which is a building block for the parity gate that will be studied in the next section. 

In order to find the density operator, we first find the normally-ordered characteristic functional for the output field operators $\hat E'_S(t)$ and $\hat E'_P(t)$, i.e., 
\begin{equation}
\label{chct}
\hspace{-.5in}
\chi_N(\zeta_S(t),\zeta_P(t)) \equiv 
\left\langle e^{\int\! {{\rm d} t\, \zeta_S(t) \hat E'^\dag_S(t)}}e^{\int\! {{\rm d} t\, \zeta_P(t) \hat E'^\dag_P(t)}}e^{-\int\! {{\rm d} t\, \zeta_P^\ast(t) \hat E'_P(t)}}e^{-\int\! {{\rm d} t\, \zeta_S^\ast(t) \hat E'_S(t)}} \right \rangle .
\end{equation}
The exact evaluation of the above functional is, in general, a tedious task. However, for operation within the slow-response regime, it is sufficient to employ the discrete-mode picture, characterized by the $\hat a_S$ and $\hat a_P$ modes, for the input field operators, in which
\begin{eqnarray}
& \hat E_S(t) = \phi(t+t_h) \hat a_S + \mbox{vacuum-state\ modes} & \\
& \hat E_P(t) = \phi(t) \hat a_P + \mbox{vacuum-state\ modes} &
\end{eqnarray}
because the slow-response condition ensures that the output field operators will then obey
\begin{eqnarray}
& \hat E'_S(t) = \phi(t+t_h) e^{i\hat \xi_S(-t_h)}e^{i\hat \mu_S(-t_h)} \hat a_S + \mbox{vacuum-state\ modes} & \\
& \hat E'_P(t) = \phi(t) e^{i\hat \xi_P(0)}e^{i\hat \mu_P(0)} \hat a_P + \mbox{vacuum-state\ modes} .&
\end{eqnarray}
The two implicit assumptions in the above equations are $t_0 = 0$ and $t_h > t_0$, where $t_0$ is the center of $|\phi(t)|$, and $t_h$ is the delay between the signal and the probe beams at which the maximum nonlinearity will be induced. 
Now, it is easy to see that if $\int\!{{\rm d} t\, \zeta_S^\ast(t) \phi(t+t_h)} = 0$ and $\int\!{{\rm d} t\, \zeta_P^\ast(t) \phi(t)} = 0$, then the averaging in equation~(\ref{chct}) is over the vacuum modes, whose normally-ordered characteristic functions are unity. Hence, the only functions for which the value of $\chi_N(\zeta_S(t),\zeta_P(t))$ is nontrivial are $\zeta_S(t) = \zeta_S \phi(t+t_h)$ and $\zeta_P(t) = \zeta_P \phi(t)$. An equivalent characteristic function can then be obtained using the discrete-mode operators
\begin{eqnarray}
&  \hat a'_S \equiv e^{i\hat \xi_S(-t_h)}e^{i\hat \mu_S(-t_h)} \hat a_S & \\
&  \hat a'_P \equiv e^{i\hat \xi_P(0)}e^{i\hat \mu_P(0)} \hat a_P , &
\end{eqnarray}
where 
\begin{eqnarray}
\label{signalph}
 \hat \mu_S(-t_h) & = &\kappa \int\!{{\rm d} \tau\, h(-t_h - \tau ) \hat E_P^\dag (\tau) \hat E_P(\tau)} \nonumber \\
 & \approx & \kappa h(-t_h )\int\!{{\rm d} \tau\,  \hat E_P^\dag(\tau)  \hat E_P(\tau)} \nonumber \\
 & = & \kappa h(-t_h) \hat a_P^\dag \hat a_P = 0.
\end{eqnarray}
and similarly, 
\begin{equation}
\label{probeph}
 \hat \mu_P(0) \approx \theta \hat a_S^\dag \hat a_S, \quad\mbox{where \ $\theta \equiv \kappa h(t_h)$.} 
\end{equation}
In both equations~(\ref{signalph}) and (\ref{probeph}) we have dropped number-operator terms for modes that are in their vacuum states.

Equation (\ref{signalph}) reflects the fact that the probe beam does not phase shift the signal pulse, because it lags the signal pulse in time and the cross-Kerr effect is causal. On the other hand, the signal pulse induces a phase shift $\theta$ on the probe beam, as shown in equation~(\ref{probeph}).  Our new characteristic function will then read
\begin{eqnarray}
\label{CHavg}
\chi_N(\zeta_S,\zeta_P)& \equiv & \langle e^{\zeta_S \hat a'^\dag_S } e^{\zeta_P \hat a'^\dag_P} e^{-\zeta_P^\ast \hat a'_P } e^{-\zeta_S^\ast \hat a'_S} \rangle \nonumber \\
  & = & |\alpha|^2 \langle e^{-2i {\rm Im} \{ \alpha_P \zeta_P^\ast \exp[i \hat \xi_P(0)] \} } \rangle \nonumber \\
& + & |\beta|^2 (1-|\zeta_S|^2) \langle e^{-2i {\rm Im} \{ \alpha_P \zeta_P^\ast \exp[i \hat \xi_P(0)] \exp[i \theta] \} } \rangle \nonumber \\
& + & \alpha \beta^\ast \zeta_S \langle e^{-i\hat \xi_S(-t_h)} e^{-2i {\rm Im} \{ \alpha_P \zeta_P^\ast \exp[i \hat \xi_P(0)] \}} \rangle \nonumber \\
& - & \alpha^\ast \beta \zeta_S^\ast \langle e^{-2i {\rm Im} \{ \alpha_P \zeta_P^\ast \exp[i \hat \xi_P(0)] \}} e^{i\hat \xi_S(-t_h)} \rangle,
\end{eqnarray}
where the averaging is taken over the phase-noise terms $\hat \xi_S(-t_h)$ and $\hat \xi_P(0)$. 
In order to perform the averaging in equation~(\ref{CHavg}), we will rewrite the phase-noise terms in a normally-ordered form by introducing an annihilation operator $\hat b$, whose quadrature components are the Hermitian operators $\hat b_1 \equiv {\rm Re}\{\hat b\} = \hat \xi_S(-t_h) / \sqrt{2\theta} $ and $\hat b_2 \equiv {\rm Im}\{ \hat b\} = \hat \xi_P(0) / \sqrt{2\theta}$. Because $[\hat \xi_S(-t_h),\hat \xi_P(0)] = i \theta$, we see that $\hat b$ satisfies $[\hat b, \hat b^\dag] = 1$. Moreover, because $\hat \xi_S$ and $\hat \xi_P$ are in thermal states, that will also be the case for $\hat b$.  Thus, with $N \equiv \langle \hat b^\dag \hat b \rangle $, we get
\begin{eqnarray}
\label{ph_var}
\sigma^2 & \equiv & \langle \hat \xi_S^2(-t_h) \rangle = \langle \hat \xi_P^2(0) \rangle \nonumber \\
& = & (2N+1)\theta/2  \nonumber \\
& = & \kappa \int\!{\frac {{\rm d} \Omega}{2\pi}\,H_i(\Omega) \coth[\hbar \Omega/ (2k_BT)]}.
\end{eqnarray}
In our new formalism, we can replace $\langle f(\hat b ^ \dag) g(\hat b) \rangle$ with $\langle f( \beta ^ \ast) g( \beta) \rangle_\beta$, where $\langle \cdot \rangle_\beta$ denotes statistical averaging over $\beta \equiv \beta_1 + i \beta_2$, which is a classical, zero-mean, isotropic, complex-valued Gaussian random variable with variance $N$.  Using this fact, we then obtain for arbitrary complex parameters $\lambda$ and $\eta$
\begin{equation}
\label{phnavg}
\langle e^{\lambda i \hat \xi_S(-t_h)}e^{\eta i \hat \xi_P(0)} \rangle =
 e^{ -\lambda\eta i \theta/2} \langle e^{\lambda i \xi_S} e^{\eta i  \xi_P} \rangle_{\xi_S,\xi_P}, 
\end{equation}
where  $\xi_S$ and  $\xi_P$ are independent, identically distributed, zero-mean, real-valued Gaussian variables with common variance $\sigma^2$.
An interesting observation is that for nonzero values of $\lambda$ and $\eta$, there exists a nontrivial phase shift $e^{ -\lambda\eta i \theta/2}$, even when the phase-noise variance $\sigma^2$ approaches zero. This is a consequence of the non-commuting nature of the phase-noise operators $\hat \xi_S(-t_h)$ and $\hat \xi_P(0)$.

The above formula makes it possible to handle the phase-noise averaging in equation~(\ref{CHavg}). 
In particular, it is easy to verify that the following density operator corresponds to the characteristic function $\chi_N(\zeta_S,\zeta_P)$
\begin{eqnarray}
\label{SPrho}
\hat \rho_{SP}(\alpha,\beta) & = & \langle \,  |\alpha|^2 |0\rangle_{a'_S}\langle 0| \otimes |\alpha_P e^{i \xi_P}\rangle_{a'_P} \langle\alpha_P e^{i \xi_P}| \nonumber \\
& + & |\beta|^2 |1\rangle_{a'_S}\langle 1| \otimes |\alpha_P e^{i (\xi_P+\theta)}\rangle_{a'_P} \langle\alpha_P e^{i (\xi_P + \theta)}| \nonumber \\
& + & \alpha^\ast \beta e^{i\xi_S}|1\rangle_{a'_S}\langle 0| \otimes |\alpha_P e^{i (\xi_P+\theta/2)}\rangle_{a'_P} \langle\alpha_P e^{i (\xi_P + \theta/2)}| \nonumber \\
& + & \alpha \beta^\ast e^{-i\xi_S}|0\rangle_{a'_S}\langle 1| \otimes |\alpha_P e^{i (\xi_P+\theta/2)}\rangle_{a'_P} \langle\alpha_P e^{i (\xi_P + \theta/2)}| \, \rangle_{\xi_S,\xi_P}.
\end{eqnarray}
The final averaging over the classical variables will be applied later, when we calculate the fidelity of the parity gate. It is important to note that the density operator predicted by a single-mode XPM theory, as used in \cite{weak} does \em not\/\rm\ coincide with the above density operator evaluated at zero phase-noise variance, i.e., at $\xi_S = \xi_P = 0$.  Whereas the first two terms in equation~(\ref{SPrho}) also appear in the density operator associated with the single-mode treatment, this is not the case for the last two terms. The reason for this difference is  the extra phase term that appeared in equation~(\ref{phnavg}) because of the non-commuting nature of the signal and probe phase-noise operators.

\section{Parity-gate fidelity analysis}
\label{Sec5parity}

\begin{figure}
\centering
\includegraphics [width=\linewidth] {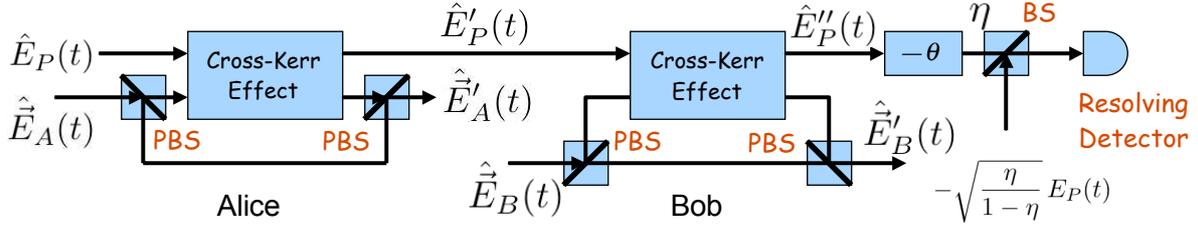}
\caption{\label{Shapiro_Razavi_fig2}
An optics-based distributed parity gate that uses a weak cross-Kerr nonlinearity. Alice and Bob encode their qubits (single photons) in the horizontal and vertical polarizations of single-photon pulses. The polarizing beam splitters (PBS) guide Alice's horizontal polarization and Bob's vertical polarization to a Kerr medium in which they interact with a coherent-state probe beam. Each such interaction can induce a phase shift $\theta$ on the probe beam. The $-\theta$ phase shifter  deterministically changes the probe's phase, and the final beam splitter (BS), with near-unity transmissivity $\eta$, models a displacement operator that is needed prior to measurement with a photon-number resolving detector.}
\end{figure}

A parity gate accepts two input qubits, in the general form $|\psi_{\rm in}\rangle = \beta_0|11\rangle_{AB} + \beta_1 |10\rangle_{AB} + \beta_2|01\rangle_{AB} + \beta_3|00 \rangle_{AB}$, and it provides a classical outcome that heralds whether the output is the even-parity state $ \beta_0|11\rangle_{AB} + \beta_3|00 \rangle_{AB}$ or the odd-parity state $ \beta_1|10\rangle_{AB} + \beta_2|01\rangle_{AB}$.  Munro, Nemoto, and Spiller \cite{weak} cascaded two single-photon/coherent-state XPM interactions to produce the all-optical parity gate shown in figure~\ref{Shapiro_Razavi_fig2}, from which an all-optical {\sc cnot} can be constructed.   In this scheme, Alice ($A$) and Bob ($B$) encode their qubits in the horizontal ($H$) and vertical ($V$) polarizations of their respective single-photon pulses, but only Alice's $H$ polarization and Bob's $V$ polarization  interact with the coherent-state probe.\footnote{Figure~2 implicitly assumes that the cross-Kerr effect is polarization independent, so that the same probe-beam polarization can undergo XPM with Alice's horizontal polarization and Bob's vertical polarization.  This assumption entails no loss of generality, because a wave plate can be inserted in the probe-beam path between the two cross-Kerr interactions to permit the use of polarization-selective XPM in realizing the distributed parity  gate.}  A single photon on Alice's $H$ mode will impart a weak phase shift $\theta \ll 1$ to the probe beam, as will a single photon on Bob's $V$ mode.  Munro, Nemoto, and Spiller \cite{weak} used a single-mode treatment of XPM to show that it should be possible to distinguish between even and odd parity states with high fidelity.  In this section we will use continuous-time XPM theory, in the slow-response regime, to provide a more accurate fidelity analysis for their parity gate.  Our conclusion about the viability of the figure~\ref{Shapiro_Razavi_fig2} gate will be rather different from theirs.

Before delving into our continuous-time fidelity analysis, it is germane to reprise the single-mode description of the parity gate, as presented in \cite{weak}. Suppose that the bus (probe) beam is in a coherent state $|\alpha_P\rangle_P$, and Alice and Bob are initially in the state $\beta_0|HH\rangle_{AB} + \beta_1 |HV\rangle_{AB} + \beta_2|VH\rangle_{AB} + \beta_3|VV \rangle_{AB}$. Each of these terms induces a different phase shift on the coherent mode.  For instance, when Alice and Bob are in the state $|HV\rangle_{AB}$, they impose a $2\theta$ phase shift on the probe beam so that its state becomes $|\alpha_P e ^{2i\theta}\rangle_P$. After the post-XPM $-\theta$ phase shift and the $-\alpha_P$ field displacement provided by  injecting the coherent state $-\sqrt{\eta/(1-\eta)}\,\alpha_P$ at the highly-transmitting ($1-\eta \ll 1$) beam splitter of figure~\ref{Shapiro_Razavi_fig2}, the probe is left in the coherent state $|\alpha_P (e ^{i\theta}-1)\rangle_P$.  Similar calculations for all other input terms leads to the following Alice-Bob-probe output state:
\begin{eqnarray}
|\psi_{\rm out}\rangle & = &(\beta_0|HH\rangle_{AB}  + \beta_3|VV \rangle_{AB}) |0\rangle_P \nonumber \\
& + & \beta_1 |HV\rangle_{AB}|\alpha_P(e^{i \theta} -1 )\rangle_P + \beta_2|VH\rangle_{AB} |\alpha_P(e^{-i \theta} -1 )\rangle_P \nonumber \\
& \approx & (\beta_0|HH\rangle_{AB}  + \beta_3|VV \rangle_{AB}) |0\rangle_P \nonumber \\
& + & \beta_1 |HV\rangle_{AB}|i \theta \alpha_P\rangle_P + \beta_2|VH\rangle_{AB} |-i \theta \alpha_P\rangle_P, \quad\mbox{for $\theta \ll 1$.}
\end{eqnarray}
Assume that the number-resolving detector has unity quantum efficiency, and that the coherent-state strength is such that $\exp(- \theta^2 |\alpha_P|^2  ) \ll 1$. Then, when no photons are detected, we will conclude that  Alice and Bob are in the even-parity state 
\begin{equation}
\label{psi0}
|\psi_0\rangle_{AB} = \frac{\beta_0|HH\rangle_{AB}  + \beta_3|VV \rangle_{AB}}{\sqrt{|\beta_0|^2 + |\beta_3|^2}}
\end{equation}
because $|\langle 0| \pm i \theta \alpha_P  \rangle|^2 = \exp(- \theta^2 |\alpha_P|^2 )\ll 1$.
Likewise, when $n \geq 1$ photons are detected, we will conclude that Alice and Bob are in the joint state
\begin{eqnarray}
\label{psin}
|\psi_n \rangle_{AB} & = & \frac{ {}_P \langle n|\psi_{out}\rangle} {\sqrt{{\rm tr}[{}_P \langle n|\psi_{out}\rangle \langle \psi_{out}|n\rangle_P]}} \nonumber\\
& \approx & \frac { \beta_1 \,  {}_P \langle n|i \theta \alpha_P\rangle_P |HV\rangle_{AB}+ \beta_2 \, {}_P \langle n|-i \theta \alpha_P\rangle_P |VH\rangle_{AB} } {\sqrt{{\rm tr}[{}_P \langle n|\psi_{out}\rangle \langle \psi_{out}|n\rangle_P]}},  \quad\mbox{for $\theta \ll 1$} \nonumber \\
& = & \frac {\beta_1 |HV\rangle_{AB} + (-1)^n \beta_2|VH\rangle_{AB} }{\sqrt{|\beta_1|^2 + |\beta_2|^2}},
\quad\mbox{for $n \geq 1$}
\end{eqnarray}
where $|n\rangle_P$ is the probe's $n$-photon Fock state.
When $n$ is an odd integer, we can apply a $\pi$-rad phase shift to make all the $n \geq 1$ observations result in the odd-parity output state $(\beta_1 |HV\rangle_{AB} + \beta_2|VH\rangle_{AB} )/ {\sqrt{|\beta_1|^2 + |\beta_2|^2}}$.
 
A key figure of merit for the distributed parity gate is the success-probability distribution, $\{P_n\}$, i.e., the probability of being in the desired state $|\psi_n\rangle_{AB}$ when $n$ probe photons have been detected. The total success probability is then given by
\begin{eqnarray}
P_{\rm success} & \equiv & \sum_{n=0}^\infty P_n = 
\sum_{n=0}^\infty {\overline{|{}_P\langle n|{}_{AB}\langle \psi_n | \psi_{\rm out} \rangle|^2}} \nonumber \\
& = & P_{\rm even} + P_{\rm odd} ,
\end{eqnarray}
where $\overline {[\cdot]}$ denotes Bloch-sphere averaging,
\begin{equation}
\label{Pesing}
P_{\rm even} \equiv P_0 = \overline{|{}_P\langle 0|{}_{AB}\langle \psi_0 | \psi_{\rm out} \rangle|^2} 
 =  \overline{|\beta_0|^2 + |\beta_3|^2 }  = 1/2 
\end{equation}
is the average success probability for the even-parity case, and
\begin{equation}
\label{Posing}
P_{\rm odd}  \equiv  \sum_{n=1}^\infty P_n = \sum_{n=1}^\infty {\overline{|{}_P\langle n|{}_{AB}\langle \psi_n | \psi_{\rm out} \rangle|^2}} 
 =   \frac {1 - e^{- \theta^2 |\alpha_P|^2}}{2}
\end{equation}
is the average success probability for the odd-parity case.

The preceding single-mode treatment shows that $\theta |\alpha_P| \gg 1$ yields near-unity success probability, whereas $\theta |\alpha_P| \ll 1$ leads to $P_{\rm success} \approx 1/2$ indicating that it is then impossible to distinguish between the odd- and even-parity states in $|\psi_{\rm out}\rangle$. The single-mode model, however, does not account for the causality-induced phase noise that is intrinsic to continuous-time XPM theory.  To evaluate the fidelity of the parity gate, for the slow-response-regime continuous-time theory, we employ the formalism introduced in the previous section. For simplicity, we assume Alice and Bob are in a normalized tensor-product state $|\psi_A\rangle |\psi_B\rangle$, where
\begin{equation}
|\psi_A\rangle = \alpha |0\rangle_A + \beta |1\rangle_A {\rm \ \ and \ \ } |\psi_B\rangle = \alpha' |0\rangle_B + \beta' |1\rangle_B ,
\end{equation}
with
\begin{equation}
 |0\rangle_K = |{\bf 0} \rangle_K {\rm \ \ and \ \ } 
 |1\rangle_K = \int\! {{\rm d} t\, \phi(t+t_h) |1_t\rangle_K} \quad\mbox{for $K=A,B$}. 
\end{equation}
Here, $|1\rangle_K$, for $K=A,B$, represents the qubit that has nonlinear interaction with the probe beam, and, for simplicity, we have assumed that there is no propagation delay between $A$ and $B$. Hence, according to figure~\ref{Shapiro_Razavi_fig2}, the following correspondence holds between the number of excitations and the polarizations of  Alice's and Bob's qubits: 
\begin{equation}
 |0\rangle_A = |V\rangle_A , |1\rangle_A = |H\rangle_A {\rm \ \ and \ \ }
 |0\rangle_B = |H\rangle_B , |1\rangle_B = |V\rangle_B .
\end{equation}
The probe beam is in the time-shifted coherent state $|\alpha_P \phi(t) \rangle_P$ that undergoes the maximum nonlinear phase shifts from Alice's and Bob's qubits. Then, the output density operator for Alice, Bob, and the probe  can be obtained by using the density operator in equation~(\ref{SPrho}) twice; once for the Alice-probe interaction and once for the Bob-probe interaction. Note that the density operator for the state associated with field operators $\hat E'_A(t)$ and $\hat E'_P(t)$ is given exactly by equation~(\ref{SPrho}), by replacing the $S$ subscripts with $A$'s. At that point in the figure~\ref{Shapiro_Razavi_fig2} setup, the probe beam is in a superposition of four different coherent states, given by equation~(\ref{SPrho}).  For each of these coherent states we can employ equation~(\ref{SPrho}) again to obtain its corresponding output density operator. The last step is to replace $\alpha_P e^{i \gamma}$ with $\alpha_P (e^{i (\gamma-\theta)}-1)$ to incorporate the effects of the phase shifter and the displacement operation. The final density operator is then given by the following equation, in which the states are associated with the output annihilation operators $\hat a'_A = \int\!{{\rm d} t\, \phi^\ast(t+t_h) \hat E'_A(t)}$, $\hat a'_B = \int\!{{\rm d} t\, \phi^\ast(t+t_h) \hat E'_B(t)}$, and $\hat a''_P = \int\!{{\rm d} t\, \phi^\ast(t) \hat E''_P(t)}$:
\begin{eqnarray}
\label{rhoAPB}
\hspace{-.4in}
\hat \rho_{APB} (\alpha,\beta, \alpha',\beta') &=& \left \langle |\alpha|^2 |0\rangle_{AA} \langle 0 | \hat \rho_{PB}^{(00)} + |\beta|^2 |1\rangle_{AA} \langle 1 | \hat \rho_{PB}^{(11)} \right. \nonumber \\
& + & \left. \alpha^\ast \beta e^{i\xi_A} |1\rangle_{AA} \langle 0 | \hat \rho_{PB}^{(10)} + \alpha \beta ^\ast e^{-i\xi_A} |0\rangle_{AA} \langle 1 | \hat \rho_{PB}^{(01)} \right\rangle_{\xi_A,\xi_B,\xi_P}
\end{eqnarray}
where
\begin{eqnarray}
\hspace{-.4in}
\hat \rho_{PB}^{(00)} & = & |\alpha'|^2 |0\rangle_{BB} \langle 0 | \otimes \left| \alpha_P (e^{i (\xi_P-\theta)}-1) \right\rangle_{PP} \left\langle \alpha_P (e^{i (\xi_P-\theta)}-1) \right| \nonumber \\
& + & |\beta'|^2 |1\rangle_{BB} \langle 1 | \otimes \left| \alpha_P (e^{i \xi_P}-1) \right\rangle_{PP} \left\langle \alpha_P (e^{i \xi_P}-1) \right| \nonumber \\
& + & \alpha'^\ast \beta' e^{i\xi_B} |1\rangle_{BB} \langle 0 | \otimes \left| \alpha_P (e^{i (\xi_P-\theta/2)}-1) \right\rangle_{PP} \left\langle \alpha_P (e^{i (\xi_P-\theta/2)}-1) \right| \nonumber \\
& + & \alpha' \beta'^\ast e^{-i\xi_B} |0\rangle_{BB} \langle 1 | \otimes \left| \alpha_P (e^{i (\xi_P-\theta/2)}-1) \right\rangle_{PP} \left\langle \alpha_P (e^{i (\xi_P-\theta/2)}-1) \right| ,
\end{eqnarray}
\begin{eqnarray}
\hspace{-.4in}
\hat \rho_{PB}^{(11)} & = & |\alpha'|^2 |0\rangle_{BB} \langle 0 | \otimes \left| \alpha_P (e^{i \xi_P}-1) \right\rangle_{PP} \left\langle \alpha_P (e^{i \xi_P}-1) \right| \nonumber \\
& + & |\beta'|^2 |1\rangle_{BB} \langle 1 | \otimes \left| \alpha_P (e^{i (\xi_P + \theta)}-1) \right\rangle_{PP} \left\langle \alpha_P (e^{i (\xi_P + \theta)}-1) \right| \nonumber \\
& + & \alpha'^\ast \beta' e^{i\xi_B} |1\rangle_{BB} \langle 0 | \otimes \left| \alpha_P (e^{i (\xi_P+\theta/2)}-1) \right\rangle_{PP} \left\langle \alpha_P (e^{i (\xi_P+\theta/2)}-1) \right| \nonumber \\
& + & \alpha' \beta'^\ast e^{-i\xi_B} |0\rangle_{BB} \langle 1 | \otimes \left| \alpha_P (e^{i (\xi_P+\theta/2)}-1) \right\rangle_{PP} \left\langle \alpha_P (e^{i (\xi_P+\theta/2)}-1) \right| ,
\end{eqnarray}
\begin{eqnarray}
\hspace{-.4in}
\hat\rho_{PB}^{(10)} = \hat \rho_{PB}^{(01)} & = & |\alpha'|^2 |0\rangle_{BB} \langle 0 | \otimes \left| \alpha_P (e^{i (\xi_P-\theta/2)}-1) \right\rangle_{PP} \left\langle \alpha_P (e^{i (\xi_P - \theta/2)}-1) \right| \nonumber \\
& + & |\beta'|^2 |1\rangle_{BB} \langle 1 | \otimes \left| \alpha_P (e^{i (\xi_P + \theta/2)}-1) \right\rangle_{PP} \left\langle \alpha_P (e^{i (\xi_P + \theta/2)}-1) \right| \nonumber \\
& + & \alpha'^\ast \beta' e^{i\xi_B} |1\rangle_{BB} \langle 0 | \otimes \left| \alpha_P (e^{i \xi_P}-1) \right\rangle_{PP} \left\langle \alpha_P (e^{i \xi_P}-1) \right| \nonumber \\
& + & \alpha' \beta'^\ast e^{-i\xi_B} |0\rangle_{BB} \langle 1 | \otimes \left| \alpha_P (e^{i \xi_P}-1) \right\rangle_{PP} \left\langle \alpha_P (e^{i \xi_P}-1) \right| ,
\end{eqnarray}
and the average in equation~(\ref{rhoAPB}) is taken over the phase noise terms $\xi_A$, $\xi_B$, and $\xi_P$, which are statistically independent zero-mean Gaussian random variables with respective variances $\sigma_A^2 = \sigma_B^2 =  \sigma^2$ and $\sigma_P^2 = 2 \sigma^2$.

Using the above density operator, we can obtain the average success probabilities for the even- and odd-parity cases as follows
\begin{eqnarray}
\label{Peven}
P_{\rm even} 
& = & f_0(\alpha_P,0)
 { \left[ C + D e^{-\sigma_P^2/2} \right] } \nonumber \\
& \cong & \frac{1} {\sqrt{1 + 2 \sigma_P^2 |\alpha_P|^2 }}
(C+De^{-\sigma_P^2/2}), \quad\mbox{when \ $\sigma_P  \ll 1$}
\end{eqnarray}
and
\begin{eqnarray}
\label{Podd}
P_{\rm odd} 
& = & C \left[1- f_0(\alpha_P,\theta)\right] 
 +  D e^{-\sigma_P^2/2} 
\left[ f_0(\sqrt{2}\alpha_P,0)-f_0(\alpha_P,0) \right] \nonumber\\
&\approx&
C \left[1- \frac{e^{- \theta^2 |\alpha_P|^2 /(1 + 2  \sigma_P^2 |\alpha_P|^2)}} {\sqrt{1+ 2  \sigma_P^2 |\alpha_P |^2 }}\right] \nonumber \\
& + & D e^{-\sigma_P^2/2} 
\left[  \frac{1} {\sqrt{1 + 4  \sigma_P^2 |\alpha_P|^2 }} - \frac{1} {\sqrt{1 + 2  \sigma_P^2 |\alpha_P |^2 }} \right], \quad\mbox{$\sigma_P + \theta \ll 1$},
\end{eqnarray}
where 
\begin{eqnarray}
\label{fnalpha}
f_0(\alpha_P,\theta) & \equiv & \left\langle \left| {}_P \langle 0|\alpha_P(e^{i(\xi_P+\theta)}-1) \rangle_P \right|^2 \right \rangle_{\xi_P} \nonumber \\
& \approx & \left\langle  e^{-|\alpha_P(\xi_P+\theta)|^2} \right \rangle_{\xi_P}, \quad\mbox{$\sigma_P+\theta \ll 1$,} \nonumber \\
& =& \frac{ e^{-\theta^2 |\alpha_P |^2 /(1+2 \sigma_P^2|\alpha_P |^2 )}} {\sqrt{1+ 2 \sigma_P^2|\alpha_P |^2 }} 
\end{eqnarray}
and
\begin{equation}
 C \equiv \overline{ \left[ \frac{|\alpha \beta'|^4 + |\alpha' \beta|^4 } {|\alpha \beta'|^2 + |\alpha' \beta|^2} \right] } \cong 0.383 {\rm \ \ and \ \ }
 D \equiv \overline{ \left[ \frac {2 |\alpha \beta \alpha' \beta'|^2 }{|\alpha \beta'|^2 + |\alpha' \beta|^2} \right] } \cong 0.117.
\end{equation}
In this analysis we have assumed that Alice's and Bob's initial states are independent and uniformly distributed over their respective Bloch spheres.

The parity gate's total success probability, $P_{\rm success} = P_{\rm even} + P_{\rm odd}$, measures how often it performs the required parity-separation task. Ideally,  $P_{\rm success} \approx 1$, something that only occurs when both of its constituents, $P_{\rm even}$ and $P_{\rm odd}$, approach their maximum values of  $1/2$.  From equation~(\ref{Peven}), we see that ideal $P_{\rm even}$ behavior occurs when $ \sigma^2_P |\alpha_P|^2\ll 1$ and $\sigma^2_P \ll 1$. 
This contrasts with the single-mode theory for $P_{\rm even}$, from equation~(\ref{Pesing}), in  which $P_{\rm even}$ is always $1/2$. 
From equation~(\ref{Podd}) we see that the $D$ term in $P_{\rm odd}$ is negative but it vanishes when either $ \sigma^2_P |\alpha_P|^2\ll 1$ or $ \sigma^2_P |\alpha_P|^2 \gg 1$.  We also have that the $C$ term in $P_{\rm odd}$ is positive, achieving its maximum value of 0.38 when $ \theta^2 |\alpha_P|^2 \gg 1$.   This contrasts with the single-mode theory, from equation~(\ref{Posing}), in which 
$P_{\rm odd}$ approaches $1/2$ for $ \theta^2 |\alpha_P|^2 \gg 1$. The reason for this difference is likely what we cited in equations (\ref{phnavg}) and (\ref{SPrho}) regarding the extra phase terms that are due to the non-commuting phase-noise operators.
Overall, based on our continuous-time theory for the cross-Kerr effect, it seems that for the optimum performance of the gate we have to satisfy the following three conditions simultaneously:
\begin{equation}
 \theta^2 |\alpha_P|^2 \gg 1, \, \,  \sigma_P^2 |\alpha_P|^2 \ll 1, \, \,\sigma_P^2 \ll 1.
\end{equation}
In the next section, we numerically evaluate $P_{\rm even}$ and $P_{\rm odd}$---first with arbitrary choices of $\sigma_P^2$ and $\theta|\alpha_P|$ and then with a two-pole response function---to investigate whether these constraints can be satisfied.

\section{Numerical results}
\label{Sec5numeric}

\begin{figure}
\centering
\includegraphics [height=2.5in] {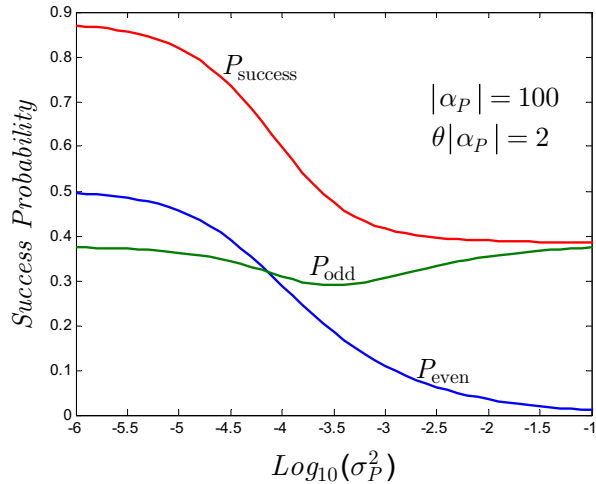}
\caption{\label{Shapiro_Razavi_fig3}
Success probabilities $P_{\rm even}$, $P_{\rm odd}$, and $P_{\rm success}$ versus phase noise variance for  $|\alpha_P| =100$ and $\theta= 20 \,$mRad.}
\end{figure}

\begin{figure}
\centering
\includegraphics [height=2.5in] {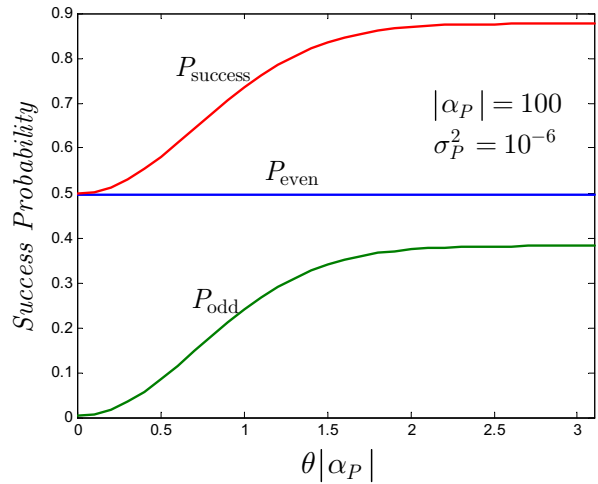}
\caption{\label{Shapiro_Razavi_fig4}
Success probabilities $P_{\rm even}$, $P_{\rm odd}$, and $P_{\rm success}$ versus $ \theta |\alpha_P|$ for  $\sigma_P^2 =10^{-6}$ and $|\alpha_P|=100$.}
\end{figure}

\begin{figure}
\centering
\includegraphics [height=3in] {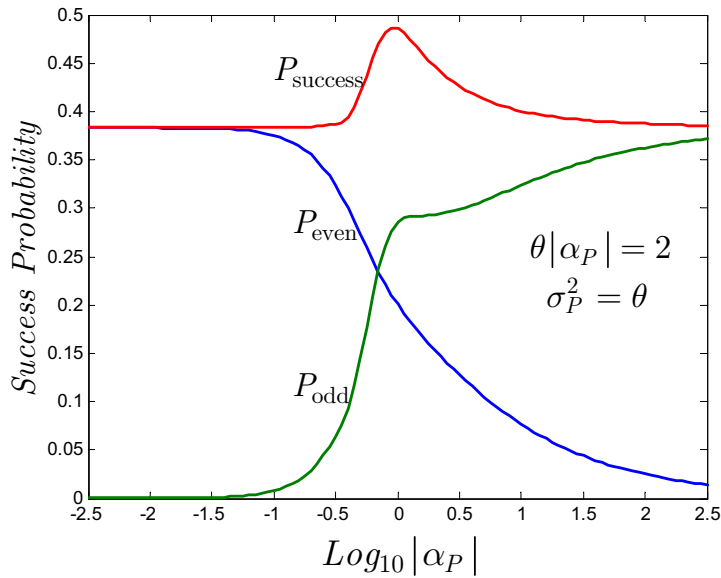}
\caption{\label{Shapiro_Razavi_fig5}
Success probabilities $P_{\rm even}$, $P_{\rm odd}$, and $P_{\rm success}$ versus the probe parameter $|\alpha_P|$ using the two-pole response function for the medium at $T=0\,K$ and $\gamma_0 \rightarrow 0$.}
\end{figure}

 In figures~\ref{Shapiro_Razavi_fig3} and \ref{Shapiro_Razavi_fig4}, we have plotted the success probabilities $P_{\rm even}$, $P_{\rm odd}$, and $P_{\rm success}$ versus  $\sigma_P^2$ and $ \theta |\alpha_P|$, respectively. Here, we have assumed that it is possible to change the phase variance $\sigma_P^2$ without affecting $\theta $, even though both are related to the cross-Kerr medium's response function. Figure~\ref{Shapiro_Razavi_fig3} shows that $P_{\rm even}$ is very sensitive to the phase noise, dropping significantly if $\sigma_P^2 |\alpha_P|^2 > 1$. On the other hand, $P_{\rm odd}$ only drops slightly for moderate values of $\sigma_P^2$, and it regains its maximum value at high values of $\sigma_P^2$. Its maximum, however, is bounded below 1/2 to the value of $C=0.38$. In figure~\ref{Shapiro_Razavi_fig4}, it can be seen that  $P_{\rm even}$ is independent of $\theta$, whereas $P_{\rm odd}$  requires $ \theta |\alpha_P| > \pi/2$ to achieve reasonable performance. That observation justifies the use of $ \theta |\alpha_P| = 2$ in figures~\ref{Shapiro_Razavi_fig3} and \ref{Shapiro_Razavi_fig5}. 

In figures~\ref{Shapiro_Razavi_fig3} and \ref{Shapiro_Razavi_fig4}, we assumed that we could independently vary $\theta$ and $\sigma_P^2$ in order to get a sufficiently large value for $\theta^2/\sigma^2_P$.  Such is not likely to be possible in our  cross-Kerr effect model because both $\sigma_P^2$ and $\theta=\kappa h(t_h)$ are functions of $h(t)$. In order to increase the maximum of $h(t)$ while keeping its area constant, we need to make it narrower. That however increases the bandwidth of $H(\Omega)$, which may increase $\sigma_P^2$. On the other hand, $\sigma_P^2 \propto {\kappa}$ and $\theta \propto {\kappa}$. Therefore, in order to satisfy both $ \theta^2 |\alpha_P|^2 > 1$ and $ \sigma_P^2 |\alpha_P|^2 \ll 1$ conditions, one needs a high ratio of $\theta ^2/ \sigma_P^2$, or equivalently, a high value for $\kappa$. This is not a desired behavior for a system that was designed to operate in the weak-nonlinearity regime. 

To clarify the above points, let's consider a concrete example. Suppose $h(t)$ is the two-pole response function associated with the two-pole frequency response 
\begin{equation}
H(\Omega)= \int\! {{\rm d} t\, h(t) e^{i \Omega t} } = \frac {\Omega_0^2}{\Omega_0^2 - \Omega^2 - i \Omega \gamma_0},
\end{equation}
where $\Omega_0$ is the vibrational resonance  frequency of the medium and $\gamma_0$ is a damping rate. This function is a good fit to the actual Raman-gain spectrum of silica fibers in the underdamped single resonance regime $0 < \gamma_0 /2 < \Omega_0$, \cite{agrawal}, in which
\begin{equation}
h(t) = \frac {\Omega_0^2 e^{-\gamma_0 t/2} \sin \left( \sqrt{\Omega_0^2 - \gamma_0^2/4} \, t \right)}{\sqrt{\Omega_0^2 - \gamma_0^2/4}}, \quad\mbox{for $t \geq 0$.}
\end{equation}
Previous work using this response function in the continuous-time XPM model  has shown \cite{jeff_kerr} that the best system performance occurs  when $T=0\,K$ and $\gamma_0 \rightarrow 0$, in which case, we have
\begin{equation}
\sigma_P^2 = (2\kappa/ \pi) \int_0^\infty\! {{\rm d} \Omega\, H_i(\Omega)} = \kappa \Omega_0. 
\end{equation}
For $\gamma_0 \rightarrow 0$, we have $h(t_h) = \Omega_0$, from which it follows that $\sigma_P^2 = \theta$, i.e., higher nonlinearity comes at the price of more phase noise.

Figure~\ref{Shapiro_Razavi_fig5} plots the success probabilities $P_{\rm even}$, $P_{\rm odd}$, and $P_{\rm success}$ versus $|\alpha_P|$ for the underdamped two-pole response at $T=0\,$K with  $ \theta |\alpha_P| = 2$.  It can be seen that optimum system performance is achieved at a low value of $|\alpha_P|$, for which $P_{\rm success}$ is only about $0.49$. This agrees with what we observed before. For high values of $|\alpha_P|$, we have that $\sigma_P^2 |\alpha_P|^2= {\theta} |\alpha_P|^2  = {2 | \alpha_P|}$ also has a high value, which, in turn, degrades the even-parity performance. For low values of $|\alpha_P|$, $P_{\rm odd} \rightarrow 0$ because $f_0(\alpha_P,\theta) \rightarrow 1$ in equation~(\ref{Podd}). 
 
\section{Discussion and conclusion}
\label{Sec5conc}

The results shown in figure~\ref{Shapiro_Razavi_fig5} demonstrate that our causal, non-instantaneous model for the cross-Kerr effect precludes the distributed parity gate's achieving high success-probability operation. One should also bear in mind that the fidelity we have derived is for the ideal case of no loss, no dispersion, and no SPM. Each of these effects can by itself significantly degrade system performance. Moreover, the above performance has been achieved under the slow-response conditions, which makes its practical application to fiber-based XPM highly questionable. However, there are still two other issues that should be addressed before coming to a definite conclusion about the parity gate's feasibility. First, within the range of validity of our model, we may still be able to do better if we choose an optimum response function for our medium. The material's normalized response function $h(t)$ must be causal and satisfy the following conditions 
\begin{equation}
 \int_0^\infty\! {{\rm d} t\, h(t) } = 1 {\rm \ \ and\ \ }
 H_i(\Omega) \geq 0 \quad\mbox{for \, $\Omega > 0$} .
\end{equation}
The above constraints define a convex set of functions. In this set, we are interested in finding the function $h(t)$ that maximizes the success probability of the gate. An easier problem is to find the function $h(t)$ that maximizes the ratio $\theta / \sigma_P$. Both these problems are analytically, as well as numerically, difficult. Even if we knew such a function, it would be difficult to find a material with the desired response function. Nevertheless, it would be interesting to find the ultimate possible performance of the parity gate using our continuous-time model for the cross-Kerr effect.

The second issue is the applicability of our model to atomic systems, possibly under electromagnetically-induced transparency  conditions \cite{schmidt},  illuminated by single photons. There are several proposals that use such systems to provide an effective cross-Kerr nonlinearity for single photons \cite{kimble_phase, Beau, Tombesi, Sanders}. Our model may or may not be applicable to such scenarios. In our model, we translate what we expect to occur classically in a pure cross-Kerr medium into quantum field-operator language, and then  we make it self-consistent by introducing phase noise operators with appropriate commutators. This is not necessarily what happens when a photon interacts with a single atom or a small ensemble of atoms. In these atom-interaction scenarios we expect to get some pulse-shape broadening, but how to relate this effect  to our medium's response function is yet to be investigated, although the work reported in \cite{Hofmann1, Koshino, Hofmann3} may be of value in this regard.   Moreover, if the physical reason behind the phase noise and the non-instantaneous response function is molecular vibrations, it is logical to ask how much vibrational noise a single photon may encounter when it interacts with a small number of atoms. In other words, if we don't expect that atomic vibrations are at all significant at the single-photon level, do we need to worry about phase noise or not? These concerns prevent us from immediately generalizing our assessment of the parity gate to all its possible  implementations. Nevertheless, a full field-operator treatment of the nonlinear behavior of atomic systems in response to single-photon pulses needs to be developed to properly answer the preceding questions, as the analysis we have presented for continuous-time XPM suggests that initial assessments of XPM-based two-qubit optical gates may be overly (perhaps even wildly) optimistic.  

\ack 
This work was supported by the MIT-HP Alliance.


\section*{References}
\begin{thebibliography}{10}

\bibitem{weak}
Munro W J, Nemoto K and Spiller T P 
2005 
{\em New J. Phys.} {\bf 7} 137

\bibitem{KLM}
Knill E, Laflamme R and Milburn G J 2001
{\em Nature} {\bf  409} pp~46--52

\bibitem{Pittman} Pittman T B,  Fitch M J, Jacobs B C and Franson J D
2003 {\em Phys. Rev. A} {\bf 68} 032316

\bibitem{schmidt}
Schmidt H and Imamoglu A  1996 {\em Opt. Lett.}  {\bf 21} pp~1936--8

\bibitem{haus}
Boivin L, K\"{a}rtner F X and Haus H A 1994
{\em Phys. Rev. Lett.} {\bf 73} pp~240--243

\bibitem{jeff_kerr}
Shapiro J H 2006 {\em Phys. Rev. A} {\bf 73} 062305

\bibitem{ike}
Chuang I L and Yamamoto Y 1995
{\em Phys. Rev. A}  {\bf 52} pp~3489--96

\bibitem{yuenjeff1}
Yuen H P  and Shapiro J H 1978
{\em IEEE Trans. Inform. Theory} {\bf IT-24} pp~657--68

\bibitem{Loudon}
Blow K J, Loudon R and Phoenix S J D 1990
{\em Phys. Rev. A} {\bf 42} pp~4102--14

\bibitem{agrawal}
Agrawal G P 2001 
{\em Nonlinear Fiber Optics, 3rd edition} 
(Academic Press: San Diego)

\bibitem{kimble_phase}
Turchette Q A, Hood C J, Lange W, Mabuchi H and Kimble H J 1995
{\em Phys. Rev. Lett.} {\bf 75} pp~4710--3

\bibitem{Beau}
Beausoleil R G, Munro W J, Rodrigues D A and Spiller T P 2004
{\em J. Modern Opt.} {\bf 51} pp~2441--8

\bibitem{Tombesi}
Ottaviani C, Rebi\'c S, Vitali D and Tombesi P 2006
{\em Phys. Rev. A} {\bf 73} 010301

\bibitem{Sanders}
Wang Z-B, Marzlin K-P and Sanders B C 2006
{\em Phys. Rev. Lett.} {\bf 97} 063901

\bibitem{Hofmann1}
Kojima K, Hofmann H F, Takeuchi S and Sasaki K 2003
{\em Phys. Rev. A} {\bf 68} 013803

\bibitem{Koshino}
Koshino K and Ishihara H 2004 
{\em Phys. Rev. A} {\bf 70} 013806

\bibitem{Hofmann3}
Kojima K, Hofmann H F, Takeuchi S and Sasaki K 2004
{\em Phys. Rev. A} {\bf 70} 013810



\end{thebibliography}
\end{document}